\documentclass[prd,aps,showpacs,preprintnumbers,amssymb]{revtex4}
\usepackage{axodraw}
\usepackage{color}
\usepackage{epsf}

\begin{document}
\title{%
\hfill{\normalsize\vbox{%
\hbox{}
 }}\\
{Confinement and chiral phase transitions in QCD and superQCD}}

\author{Renata Jora
$^{\it \bf a}$~\footnote[2]{Email:
 rjora@theory.nipne.ro}}

\affiliation{$^{\bf \it a}$ National Institute of Physics and Nuclear Engineering PO Box MG-6, Bucharest-Magurele, Romania}

\date{\today}

\begin{abstract}
 We study confinement and chiral symmetry breaking phase transitions for QCD and supersymmetric QCD. The method is based on some justified assumption regarding the behavior of the Green function relevant for the potential between two charges or a gap equation. We determine the number of flavors at which the transitions to confinement and chiral symmetry breaking take place. We find that whereas for chiral symmetry breaking the results are in perfect agreement with those in the literature for confinement  the situation is more complex.

\end{abstract}
\maketitle

\section{Introduction}

Phase diagram of a nonabelian gauge theory at zero or finite temperature is of great interest and consequences both for studying real life QCD phenomena and for possible standard model extensions (with dynamical electroweak symmetry breaking or technicolor). Over the years a number of results regarding the behavior in terms of the number of flavors  \cite{Appelquist1}-\cite{Jora3} or dynamical studies at zero or finite temperature \cite{Pisarski1}-\cite{Pisarski2} improved our understanding of QCD like theories in the critical regime.

Supersymmetric gauge theories are a fruitful laboratory of studying phase transitions not only  from theoretical point of view but also of the possible applications. Important exact results stemming from  the holomorphicity and the enhanced symmetry of the theory were obtained. The phase diagram in terms of the number of flavors of $N=1$ supersymmetric gauge theories has been elucidated by Seiberg in a series of groundbreaking works  \cite{Seiberg1}, \cite{Seiberg2} (and the references therein) based on the power of dualities.

In this work we will revisit confinement and chiral phase transitions from a fresh point of view. Since both QCD and supersymmetric QCD are basically gauge theories it is natural to consider an approach that treats both of them on the same footing. Using the behavior of the relevant Green function and associated Feynman diagrams that characterize confinement and chiral symmetry breaking we will establish a common set of constraints to be verified by both QCD and its supersymmetric counterpart. This will allow us to determine the behavior of the theory in terms of the number of flavors. Sections II and III contain the treatment of QCD confinement and chiral symmetry breaking respectively. Sections IV and V include a similar approach applied to supersymmetric QCD. The conclusions are drawn is section VI.

\section{Confinement in QCD}

Beta function for QCD is known up to the fifth order \cite{Vermaseren1}, \cite{Vermaseren2}. For $N_f$ fermions in the fundamental representation it has the expression:
\begin{eqnarray}
\beta(g)=\frac{d g}{d\ln\mu}=-\frac{g^3}{16\pi^2}[\beta_0+\beta_1\frac{g^2}{16\pi^2}],
\label{beta546}
\end{eqnarray}
where only the first two renormalization scheme independent coefficients were retained:
\begin{eqnarray}
&&\beta_0=\frac{11}{3}N-\frac{2}{3}N_f
\nonumber\\
&&\beta_1=\frac{34}{3}N^2-\frac{N^2-1}{N}N_f-\frac{10}{3}NN_f.
\label{coeff776}
\end{eqnarray}

The fermion mass anomalous dimension is relevant and renormalization scheme independent only at one loop:
\begin{eqnarray}
\gamma_m=\frac{dm}{d\ln\mu}=-3\frac{N^2-1}{N}\frac{g^2}{16\pi^2}.
\label{anom77688}
\end{eqnarray}

The gauge theory is confining when the potential between charges  is of the form $V\approx \sigma r$ \cite{Seiberg1} where $\sigma$ is a constant and $r$ is the distance between the charges.  In the case of QCD the potential between two charges is given by the gluon propagator. Then by simple Fourier analysis of the propagator (and also as proposed in the literature \cite{Cornwall}, \cite{Doff} the gluon propagator should be of the type,
\begin{eqnarray}
G^{\mu\nu}(p)\approx g^{\mu\nu}\frac{1}{p^4},
\label{glprop768}
\end{eqnarray}
where we used Feynman gauge.

In the perturbation theory in order for  the form  in Eq. (\ref{glprop768}) to be achieved the propagator must have the exact expression:
\begin{eqnarray}
G^{\mu\nu}(p)=g^{\mu\nu}\frac{1}{p^2}\frac{M^2}{p^2}g^2,
\label{propgl8999}
\end{eqnarray}
where $M$ is the renormalization scale and  each factor of $M$ should bring a factor of $g$ the coupling constant. In principle however confinement is present in some degrees for any potential of the type $V\approx \sigma r^s$, where $s\geq1$ which translates into a propagator:
\begin{eqnarray}
G^{\mu\nu}(p)\approx g^{\mu\nu}\frac{1}{p^2}(\frac{M}{p})^kg^k,
\label{glprop76895}
\end{eqnarray}
where $k\geq 2$.

Knowing the form of the propagator we can apply the Callan Symanzik equations:
\begin{eqnarray}
[M\frac{\partial}{\partial M}+2\frac{\beta(g)}{g}+\beta(g)\frac{\partial}{\partial g}]G^{\mu\nu}(p)=0.
\label{CSeq444}
\end{eqnarray}
Here we considered a background gauge field type of renormalization where the gluon wave function renormalization is given by $\frac{\beta(g)}{g}$.  For the expression of the gluon propagator introduced in Eq. (\ref{glprop76895}) one further obtains from Eq. (\ref{CSeq444}):
\begin{eqnarray}
k+(2+k)\frac{\beta(g)}{g}=0.
\label{firstconstr6}
\end{eqnarray}

One can regard the confining theory with the gluon propagator as giving the confining potential; however there is an alternative picture  in which the gluon field is integrated out and the theory becomes of the Nambu Jona Lasinio  type with four fermion interaction as in:
\begin{eqnarray}
\frac{1}{M^2}\bar{\Psi}\gamma^{\mu}t^a\Psi\bar{\Psi}\gamma_{\mu}t^a\Psi,
\label{fourferm77688}
\end{eqnarray}
where the flavor dependence is omitted as irrelevant. Then a diagram of the type in Fig. 1 leads to the formation of the bound states and in order for this to be confining again the overall contribution must behave as $\frac{1}{p^4}$ which means that the fermion propagator must  be:
\begin{eqnarray}
G_1(p)\approx \frac{1}{\gamma^{\mu}p_{\mu}}\frac{M}{\gamma^{\rho}p_{\rho}}g.
\label{fermprop8867}
\end{eqnarray}
For the simplicity of the notation we will denote $G_1(p)\approx \frac{1}{p}\frac{M}{p}g$ thus retaining only the features that are relevant for our purposes.

\begin{center}
\SetScale{0.8}
\begin{picture}(200,150)(0,0)
\Line(0,50)(50,100)
\Line(50,100)(0,150)
\GCirc(100,100){50}{1}
\Line(150,100)(200,50)
\Line(150,100)(200,150)
\end{picture}
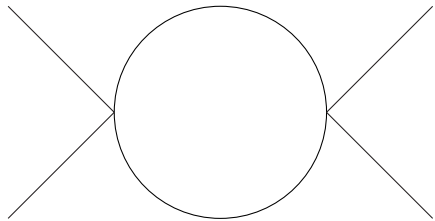
\\{\sl Fig. 1. Diagram of four particle interaction of the Nambu Jona Lasinio type. For QCD the external states and those in the loop are fermions; for supersymmetric QCD all states are scalars and in loop they are appear with derivatives}
\end{center}

Again the expression in Eq. (\ref{fermprop8867}) can be generalized to:
\begin{eqnarray}
G_1(p)=\frac{1}{p}(\frac{M}{p})^{k_1}g^{k_1},
\label{fermprop77788}
\end{eqnarray}
where $k_1\geq 1$.
We apply the Callan Symanzik equations,
\begin{eqnarray}
[M\frac{\partial}{\partial M}+2\gamma_1+\beta(g)\frac{\partial}{\partial g}]G_1(p)=0,
\label{fermcsppp99}
\end{eqnarray}
where $\gamma_1$ is the anomalous dimension of the fermion wave function and $2\gamma_1=\gamma_m$.  This leads to:
\begin{eqnarray}
k_1+\gamma_m+k_1\frac{\beta(g)}{g}=0,
\label{secondconts664554}
\end{eqnarray}

We first solve the system of equations (\ref{firstconstr6}) and (\ref{secondconts664554}) for $k=2$ and $k_1=1$. This leads to the coupling constant and the number of flavors for which the transition to the confinement phase occurs:
\begin{eqnarray}
&&\frac{g_c^2}{16\pi^2}=\frac{N}{6(N^2-1)}
\nonumber\\
&&N_{fc}=\frac{46N^4+42N^2-54}{25N^3-15N}.
\label{res5666}
\end{eqnarray}
In the large N limit we get:
\begin{eqnarray}
N_{fc}\approx \frac{46}{25}N
\label{largen00}
\end{eqnarray}
For QCD with $N=3$ $N_{fc}\approx 5.52$. This could explain why in the standard model the lightest five flavors confine forming bound states less or more stable whereas the top quarks occurs always as an elementary state.

Next we consider the case $k> 2$ and $k_1> 1$ which yields:
\begin{eqnarray}
&&\frac{g_c^2}{16\pi^2}=\frac{2k_1}{3(2+k)(N^2-1)}
\nonumber\\
&&N_{fc}=\frac{(-54k-27k^2+132k_1+66kk_1+136k_1^2)N}{4k_1(6+3k+13k_1)}+O(\frac{1}{N}).
\label{gneres66477}
\end{eqnarray}
It can be determined numerically that:
\begin{eqnarray}
N_{fc}\leq 2.8 N.
\label{limits64665}
\end{eqnarray}

\section{QCD chiral phase transition}

We claim that the breaking of a global symmetry like the chiral symmetry should be associated clearly with a phase transition and thus to a zero of the beta function. The next criterion comes from  a gap equation given by the a diagram of the type in Fig. 2 to which we associate the Green function $G_2$. This implies the existence of a fermion interaction term of the type $\frac{1}{M^2}\bar{\Psi}\Psi\bar{\Psi}\Psi$ where $M$ is the renormalization scale.

 \begin{center}
 \SetScale{0.8}
  \begin{picture}(200,150)(0,0)
 \Line(0,50)(200,50)
 \GCirc(100,100){50}{1}
 \end{picture}
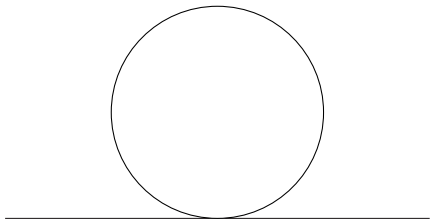
\\ {\sl Fig. 2. Diagram depicting the gap equation. For QCD all states are fermions. For supersymmetric QCD all states are scalars.}
\label{phases2}
\end{center}

Then in order for the fermion loop to contribute it must again be of the type,
\begin{eqnarray}
G_1(p)\approx\frac{1}{p}(\frac{M}{p})^{k_2}g^{k_2},
\label{resiff}
\end{eqnarray}
with $k_2\geq 1$.
Then the result of the diagram in Fig. 2 (applied to this case) which correspond to $G_2$ is:
\begin{eqnarray}
G_2\approx\int \frac{d^4p}{(2\pi)^4}\frac{1}{p}(\frac{M}{p})^{k_2}g^{k_2}\frac{1}{M^2}\approx {\rm const}Mg^{k_2}.
\label{res77564}
\end{eqnarray}
We apply to $G_2$ the Callan Symanzik equations to get:
\begin{eqnarray}
[M\frac{\partial}{\partial M}+4\gamma_1+\beta(g)\frac{\partial}{\partial g}]G_2=0.
\label{ca77688}
\end{eqnarray}
Then the two constraints  of interest become:
\begin{eqnarray}
&&\beta(g)=0
\nonumber\\
&&2\gamma_m+1=0.
\label{res6990028}
\end{eqnarray}
The second equation in (\ref{res6990028}) is a direct result of Eq. (\ref{ca77688}).

We  solve Eq. (\ref{res6990028}) which leads to:
\begin{eqnarray}
&&\frac{g^2_s}{16\pi^2}=\frac{N}{6(N^2-1)}
\nonumber\\
&&N_{fs}=\frac{2N(-33+50N^2)}{5(-3+5N^2)}.
\label{finalres775664}
\end{eqnarray}
In the large N limit one gets $N_{fs}\approx4N$.  The results in Eq. (\ref{finalres775664}) agree exactly with the studies in \cite{Appelquist1}, \cite{Appelquist2} and are in good agreement with the lattice results \cite{Kogut}.

\section{SuperQCD confinement and chiral symmetry breaking}

We consider the supersymmetric QCD holomorphic Lagrangian:
\begin{eqnarray}
{\cal L}_h=\frac{1}{16}\int d^2\theta W^a(V_h)W^a(V_h)+h.c. +\int D^4 \theta \sum_I\Phi_i^{\dagger}\exp[2V_h^i]\Phi_i.
\label{holo999}
\end{eqnarray}
Here i represents the chiral multiplet $i$ and $V_h^i=V_h^aT^a_i$ where $T^a$ are the generators of the group in the fundamental representation of the chiral multiplet.

One can make a transformation to the canonical couplings by the change of variable $V_h=g_cV_c$ \cite{Murayama}. Then the beta function in the canonical coupling is the NSVZ beta function \cite{NSVZ}:
\begin{eqnarray}
\beta(g_c)=-\frac{g_c^3}{16\pi^2}\frac{3N-N_f+N_f\gamma(g_c)}{1-N\frac{g^2_c}{8\pi^2}},
\label{betaufnc665}
\end{eqnarray}
where $\gamma(g_c)$ is the anomalous dimension of the chiral multiplet mass operator  which at one loop has the expression:
\begin{eqnarray}
\gamma(g_c)=-\frac{g_c^2}{8\pi^2}\frac{N^2-1}{N}.
\label{anom885774}
\end{eqnarray}

The nonperturbative behavior of the supersymmetric QCD and the corresponding phase structure has been analyzed and clarified by Seiberg in  \cite{Seiberg1}, \cite{Seiberg2}.  Here we shall discuss confinement and chiral symmetry breaking from a different point of view, that introduced in section II for QCD.

In supersymmetric QCD  there are gluons and gluinos in the gauge supermultiplet and fermions and scalars in the chiral supermultiplet. In order for bound states of mesons and baryons to form not only the gluon propagator must be confining but also the gluino propagator(in order to form baryons).

According to the arguments given in section II this amounts to asking for the gluino propagator to be of the type:
\begin{eqnarray}
G_{gl}(p)\approx\frac{1}{p}(\frac{M}{p})^{k_1},
\label{gluino74773663}
\end{eqnarray}
with $k_1\geq 3$. Here $k_1=3$ is the correct condition of confinement. However corrections of order $(\frac{M}{p})^3$ can come only from the chiral multiplet itself so are associated with a coupling constant $g_c^2$. Thus the complete formula for the gluino propagator should be:
\begin{eqnarray}
G_{gl}(p)\approx\frac{1}{p}(\frac{M}{p})^{k_1}g_c^{k_2},
\label{exact3442442}
\end{eqnarray}
with $k_1\geq3$ and $k_2\leq k_1-1$.

We apply the Callan Symanzik equation to $G_{gl}(p)$,
\begin{eqnarray}
[M\frac{\partial }{\partial M}+2\frac{\beta(g_c)}{g_c}+\beta(g_c)\frac{\partial}{\partial g_c}]G_{gl}(p)=0,
\label{cettr5546}
\end{eqnarray}
to get:
\begin{eqnarray}
k_1+(2+k_2)\frac{\beta(g_c)}{g_c}=0.
\label{superfirstconstr645}
\end{eqnarray}

The supersymmetric Lagrangian can be also regarded as a function only of the chiral multiplet, a generalized version including also scalars of the Nambu Jona Lasinio model.  Then by considering a virtual mass for the gluon field and integrating it out leads to four scalar interaction terms of the type:
\begin{eqnarray}
\frac{1}{M^2}(\Phi\partial^{\mu}\Phi^*)(\Phi^*\partial_{\mu}\Phi).
\label{intercs554663}
\end{eqnarray}

Again in order to get for the diagram in Fig. 1 (adjusted for this case)  a confining type of potential one would need in lowest order a scalar two point function of the type:
\begin{eqnarray}
G_s(p)\approx\frac{1}{p^2}\frac{M^2}{p^2}g_c^2,
\label{glsuper555}
\end{eqnarray}
or in the general case,
\begin{eqnarray}
G_s(p)\approx\frac{1}{p^2}(\frac{M}{p})^{k_3}g_c^{k_3},
\label{gengl996887}
\end{eqnarray}
with $k_3\geq 2$.
We apply the Callan Symanzik equations to $G_s(p)$ to get,
\begin{eqnarray}
[M\frac{\partial}{\partial M}+2\gamma_1+\beta(g_c)\frac{\partial}{\partial g_c}]G_s(p)=0,
\label{cttryyr}
\end{eqnarray}
which further yields:
\begin{eqnarray}
k_3+k_3\frac{\beta(g_c)}{g_c}+2\gamma_1=0.
\label{secconstsupersym664554}
\end{eqnarray}

We first solve the system formed out of Eqs. (\ref{superfirstconstr645}) and (\ref{secconstsupersym664554}) for $k_1=3$, $k_2=2$, $k_3=2$ to find:
\begin{eqnarray}
&&\frac{g_{cc}}{8\pi^2}=\frac{N}{2(N^2-1)}
\nonumber\\
&&N_{fc}\approx N+O(\frac{1}{N}).
\label{resultssupe55466}
\end{eqnarray}

For the general $k$'s the system of equations becomes:
\begin{eqnarray}
&&k_1+(2+k_2)\frac{\beta(g_c)}{g_c}=0
\nonumber\\
&&k_3+k_3\frac{\beta(g_c)}{g_c}+\gamma_m=0.
\label{res774663}
\end{eqnarray}
The system can besolved and leads in the large N limit to:
\begin{eqnarray}
&&\frac{g_{cc}^2}{16\pi^2}=\frac{(2-k_1+k_2)k_3}{2(2+k_2)N}
\nonumber\\
&&N_{fc}=\frac{-2k_1^2k_3+3(2+k_2)^2k_3-k_1(2+k_2)(2+k_3)}{(-2+k_1-k_2)k_3(-2-2k_3+k_1k_3--k_2-k_2k_3)}.
\label{res775999320}
\end{eqnarray}
It can be shown that $N_{fc}\leq \frac{3}{2}N$. The theory for general $k$'s is asymptotically free but the exact behavior for the full range of $N_{fc}$ cannot be specified except that there are elements of confining. For $N_f\approx N$ the theory is clearly confining fact which agrees with the Seiberg results \cite{Seiberg1}, \cite{Seiberg2}  regarding this case. However one cannot see in this approach the other confining case $N_f=N+1$ found by Seiberg.

\section{Supersymmetric QCD chiral symmetry breaking}

For chiral symmetry breaking phase transition we again will ask that $\beta(g_c)=0$.  The gap equation refers to the figure Fig. 2 for the scalars and puts no constraints on the scalar propagator. This can be of the form:
\begin{eqnarray}
G_s(p)=\frac{1}{p^2}(\frac{M}{p})^{k_4}g_c^{k_4}.
\label{respropsc65774}
\end{eqnarray}
where $k_4\geq 0$. Then the diagram in Fig. 2  behaves like (Note that here we do not need terms obtained form integrating out the gauge fields and the standard four scalar interaction in the supersymmmetric Lagrangian suffices):
\begin{eqnarray}
G_4\approx \int \frac{d^4 p}{(2\pi)^4}p^2\frac{1}{p^2}(\frac{M}{p})^{k_4}g_c^{k_4}\approx M^4.
\label{res774663}
\end{eqnarray}
We apply the Callan Symanzik equation to $G_4$ to get:
\begin{eqnarray}
4\gamma_1+4=0.
\label{finalres74663554}
\end{eqnarray}
Finally we have two constraints,
\begin{eqnarray}
&&\beta(g_c)=0
\nonumber\\
&&2\gamma_m+4=0,
\label{res788283995}
\end{eqnarray}
which lead to the clear result:
\begin{eqnarray}
&&\frac{g_{cs}^2}{16\pi^2}=\frac{N}{N^2-1}
\nonumber\\
&&N_{fs}=N.
\label{res720019}
\end{eqnarray}
This result is in complete agreement with previous work on this topic \cite{Seiberg1}, \cite{Seiberg2}.

\section{Conclusions}
In this work we studied confinement and chiral symmetry breaking phase transitions for QCD and supersymmetric QCD in terms of the number of flavors $N_f$. For chiral symmetry breaking the constraints we proposed match those in the literature for QCD and are in agreement with those for supersymmetric QCD. For QCD the critical number of flavors at which chiral symmetry breaking occurs is $N_{fs}\approx 4N$, for superQCD is $N_{fs}=N$. The type of phase transition remains undetermined and requires further investigation.

For confinement  the situation is more complex and intricate because it is always possible that the system displays an admixture of phases instead of pure confinement. Our result for the critical number of flavors for QCD is $N_{fc}\approx 1.8 N$ and does not agree with those findings in the literature which claim that chiral symmetry breaking and confinement occurs at almost the same point in the flavor space.  However for real life QCD our result has an unexpected bonus: according to it only five flavors may confine and thus confining is possible only after the top quark is integrated out.

For supersymmetric QCD the critical number of flavors at which confinement takes place is $N_{fc}\approx N$. The case $N_{fc}=N+1$ established by Seiberg \cite{Seiberg1}, \cite{Seiberg2} is possible but does not appear as distinct.   This outcome is not unexpected as our approach differs essentially from that in \cite{Seiberg1}.

The results obtained in this work  are partially in good agreement with those obtained in the literature and  different in some instances. The method proposed here may allow the study of all possible phases in the phase diagram of a gauge theory and lead to further insights into its structure and dynamics.


\begin{thebibliography}{30}
\bibitem{Appelquist1} T. Appelquist, J. Terning and L. C. R. Wijewardhana, Phys. Rev. Lett. {\bf 77}, 1214-127 (1996).
\bibitem{Appelquist2} T. Appelquist, A. Ratnaweera, J. Terning and L. C. R. Wijewardhana, Phys. Rev. D {\bf 58}, 105017 (1998).
\bibitem{Appelquist3} T. Appelquist, Z. Duan and  F. Sannino, Phys. Rev. D {\bf 61}, 125009 (2000).
\bibitem{Jora1} A. H. Fariborz and R. Jora, Mod. Phys. Lett. A {\bf 32}, no. 02, 1750008 (2016).
\bibitem{Jora2} R. Jora, Int. J. Mod. Phys. A {\bf 30}, no. 34, 1550203 (2015).
\bibitem{Jora3}  R. Jora, Phys. Rev. D {\bf 82}, 056005 (2010).
\bibitem{Pisarski1} R. D. Pisarski and V. V. Skokov, Phys. Rev. D {\bf 94}, 054008 (2016).
\bibitem{Pisarski2} R. Pisarski, V. V. Skokov and A. M. Tsvelik, arXiv:1801.08156 (2018).
\bibitem{Seiberg1} N. Seiberg, Nucl. Phys. B {\bf 435}, 129 (1995).
\bibitem{Seiberg2} N. Seiberg, Nucl. Phys. Proc. Suppl. {\bf 45} BC:1-28 (1996).
\bibitem{Vermaseren1} K. G. Chetyrkin, G. Falcioni, F. Herzog and J. A. M. Vermaseren, JHEP {\bf 1710}, 179 (2017).
\bibitem{Vermaseren2} F. Herzog, B. Ruijl, T. Ueda, J. A. M. Vermaseren and A. Vogt, JHEP {\bf 1702}, 090 (2017).
\bibitem{Cornwall} J. M. Cornwall, Phys. Rev. D {\bf 83}, 076001 (2011).
\bibitem{Doff} A. Doff, F. A. Machado and A. A. Natale, Annals Phys. {\bf 327}, 1030 (2012).
\bibitem{Kogut} E. Dagotto, J. B. Kogut and A. Kocic, Phys. Rev. Lett {\bf 62}, 1083 (1989); Nucl. Phys. B {\bf 334}, 279 (1990).
\bibitem{Murayama} N. Arkani-Hamed, H. Murayama, JHEP {\bf 0006}, 030 (2000).
\bibitem{NSVZ} V. A. Novikov, M. A. Shifman, A. I. Vainshtein and V. I.  Zakharov, Nucl. Phys. B {\bf 229}, 381 (1983).

\end{thebibliography}
\end{document}